\documentclass[12pt,journal,twocolumn]{IEEEtran}
\usepackage{graphicx}

\title{Stability of Superhydrophobic Ring \& Axle Liquid Bearings}
\author{Elliot Jenner\IEEEcompsocitemizethanks{\IEEEcompsocthanksitem Elliot Jenner was with the University of Pittsburgh, Pittsburgh, PA 15260 USA  (e-mail: elj17@pitt.edu)}, Brian D'Urso \IEEEcompsocitemizethanks{\IEEEcompsocthanksitem Brian D'Urso is with the University of Pittsburgh, Pittsburgh, PA 15260 USA (e-mail: dursobr@pitt.edu)}}

\begin{document}
\maketitle
\begin{abstract}
Friction between contacting solid surfaces is a dominant force on the micro-scale and a major consideration in the design of MEMS. Non-contact fluid bearings have been investigated as a way to mitigate this issue. Here we discuss a new design for surface tension-supported thrust bearings utilizing patterned superhydrophobic surfaces to achieve improved drag reduction. We examine sources of instability in the design, and demonstrate that it can be simply modeled and has superior stiffness as compared to other designs.
\end{abstract}


\section{Introduction}
Micro-Electro-Mechanical Systems (MEMS) are increasingly finding practical application. Friction from solid-solid contacts in bearings and hinges can lead to device failure or breakage, so practical devices have primarily used flexing components, such as cantilevers, rather than rotating shafts and gears, which are common in macro-scale devices. Frictional concerns are one of the primary obstacles to using rotating components for MEMS.\cite{ICmotor} In macroscopic devices, air bearings with an actively injected lubricating gas layer or journal bearings with a dynamically maintained lubricant film are used in high-performance applications, but engineering such intricate structures is challenging in MEMS.\cite{MITTurbine} The use of surface tension to maintain a liquid lubricant layer is an alternative that is most effective in MEMS, since surface tension is very significant in microscopic devices. Furthermore, surface tension and fluid pressure at these scales can actually be used to passively support a MEMS bearing, eliminating all solid-solid contact. In such bearings, a liquid, usually water or an aqueous solution, is used between the two solid surfaces. Variations in surface wetting (e.g. hydrophobicity) are used to pin the liquid in position. The liquid-solid contact is low friction compared to solid-solid contacts, allowing parts to slide with less stiction and lower wear.\cite{dropbearing} 

\begin{figure*}[!tbh]
\centering
	\includegraphics[width=0.95\textwidth]{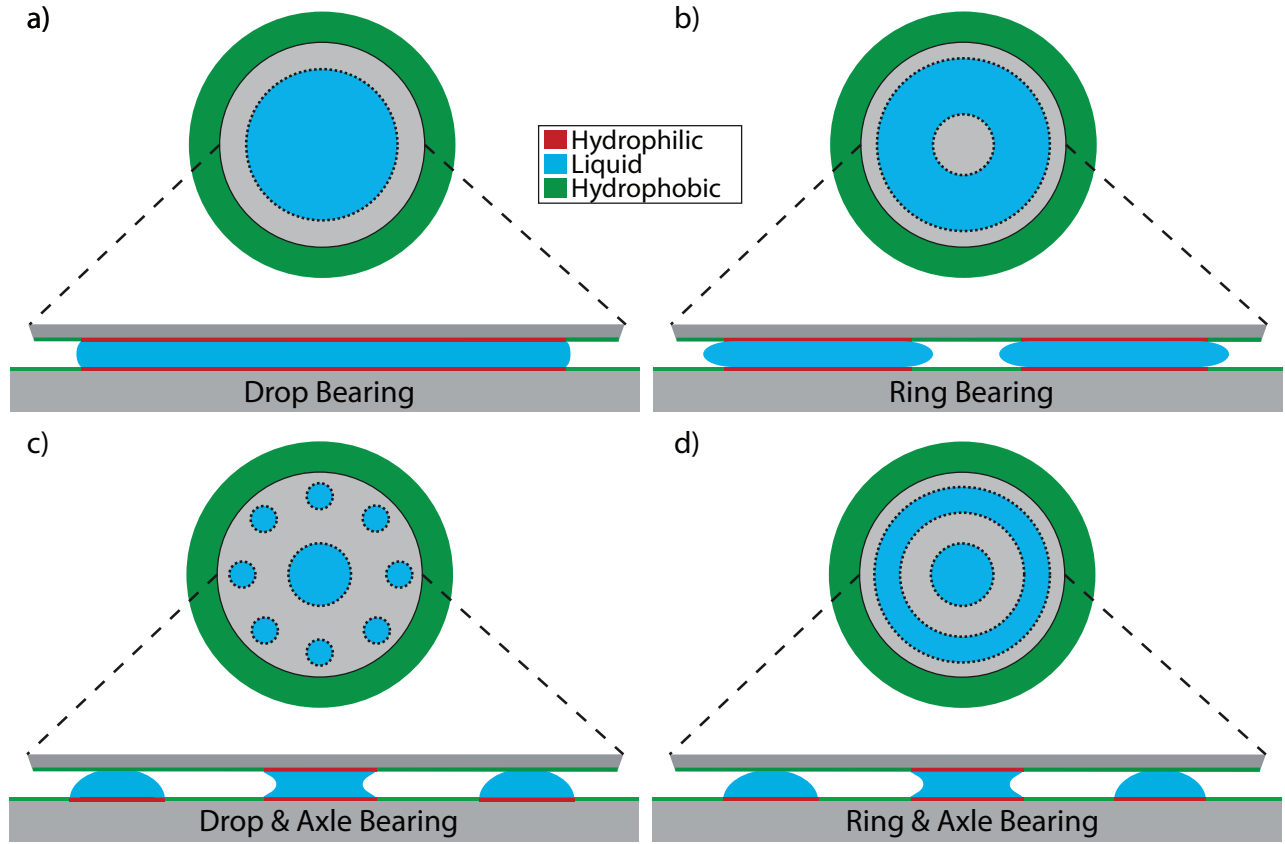}
  	\caption{Liquid bearing configurations, showing top view and cross section. Rotors are rendered in a different color to make the extent clear.}
  	\label{fig:bearing_types}
\end{figure*}

A variety of geometries for surface tension supported liquid bearings have been investigated. In the simplest case, a single drop of water trapped between identical hydrophilic pads in the center of the bearing (Fig.~\ref{fig:bearing_types}a) supports the rotating part (rotor) above the stationary substrate (stator).\cite{MEMSbearing} Wear is reduced due to the lack of solid contact and concentricity of the pads is well maintained by surface tension, but the tilt stability is poor, as there is little energetic cost to tipping the rotor even to the point of a collision between the rotor and stator. Using a ring of water reduces the hydrodynamic drag on the surfaces somewhat by reducing the wetted surface area (Fig.~\ref{fig:bearing_types}b), and has increased tilt stiffness.\cite{MEMSbearing} The tilt stiffness can be greatly increased by breaking up the wetting ring into discrete drops and making the rotor superhydrophobic, as in Fig.~\ref{fig:bearing_types}c.\cite{MEMSrotary} A superhydophobic surface is a structured surface with a water contact angle greater than 150$^\circ$.\cite{HPdef} Since the drops do not wet to the rotor surface, a central drop, which wets to both sides, is added to the center to maintain the relative positions of the surfaces. The decreased contact area between the water and solid can reduce the hydrodynamic drag, and the superhydrophobic contacts provide high lifting forces and reduce titling, but energy loss due to the remaining contact angle hysteresis of the superhydrophobic surface may dominate the sources of drag.\cite{MEMSrotary}

\begin{figure}[!tbh]
\centering
	\includegraphics[width=0.45\textwidth]{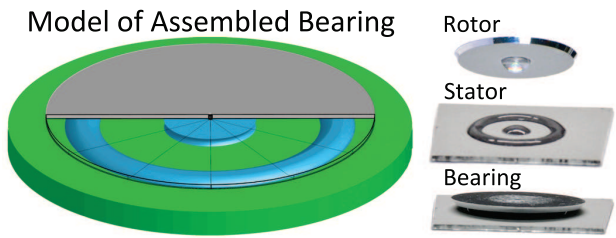}
  	\caption{A model of ring and axle liquid bearing, with photographs of a rotor, stator and assembled bearing. The model rotor is rendered as partially transparent so the liquid pattern can be seen.}
  	\label{fig:bearing}
\end{figure}

The ideal surface tension supported liquid bearing would provide high stiffness in all directions while minimizing all sources of drag. In this paper, we investigate a new ring and axle bearing design (Fig.~\ref{fig:bearing_types}d and Fig.~\ref{fig:bearing}), which combines several of the best features of other designs. The design uses a central drop, which wets hydrophilic areas on both the rotor and the stator, as an axle. The axle holds the rotor down and keeps the rotor concentric with the stator. A ring of water is wetted to a hydrophilic annulus on the stator, while the surface of the rotor in contact with the ring is superhydrophobic. The ring provides a vertical force to balance that of the axle, and stiffens the bearing with respect to tilting. Since the rotor is superhydrophobic where it contacts the ring of liquid, there may be less hydrodynamic drag than in the ring bearing, where the rotor surface in contact with the liquid is hydrophilic.\cite{HPdragreduce} Furthermore, since there is no wetting and dewetting in the system as the rotor spins, there are no hysteretic losses and thus the drag is expected to be reduced compared to the drop and axle bearing design. This new ring and axle liquid bearing design is not unconditionally stable, e.g. the ring will actually push the rotor to tilt if it is overfilled with water. Here, we examine this system and determine the parametric region of stability. 

The axle and ring each generally provide two contributions to the forces acting on the rotor: the force due to Laplace pressure, which is the result of the pressure difference across the liquid surface, and the force directly due to surface tension of the water acting on the rotor. Even in the simplest cases, calculating the stability and stiffness of the bearing analytically is not trivial, and numerical solution is more practical. We examine the bearings via numerical modeling using energy minimization.\cite{axisymmetric}

\begin{figure*}[tbh!]
  \centering
	\includegraphics[width=.95\textwidth]{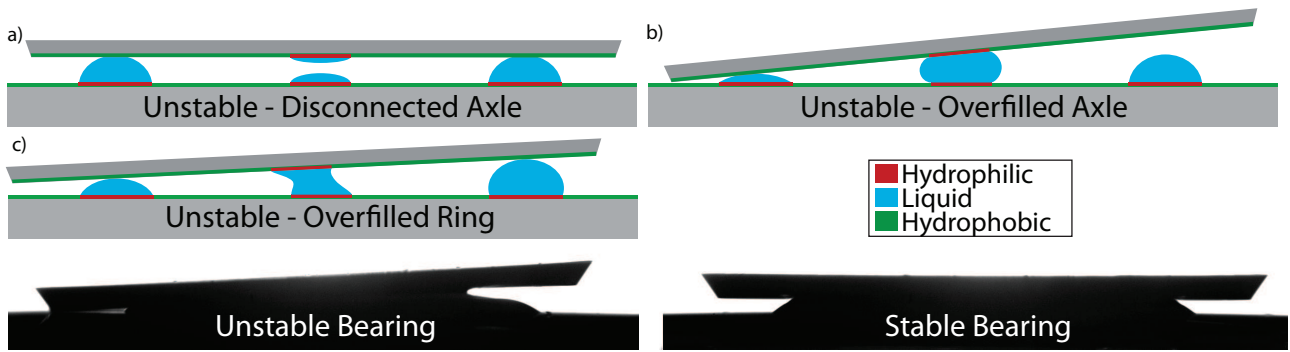}
  \caption{Illustration of possible unstable bearing states. Images of a failed and a functional bearing are shown below.}
  \label{fig:failure}
\end{figure*}

\section{Modeling}
We performed simulations of our bearings using Surface Evolver\cite{SurfaceEvolver}, an open source finite element analysis system. Surface Evolver minimizes the surface energy of the liquid subject to constraints, which include the size and contact angles of the wetting and non-wetting surfaces, and the volume of liquid in each region. The effect of gravity is also included, using experimental values for the mass of the rotor. Only the liquid surface was directly modeled; the rotor and stator were represented by boundary conditions and constraints. Typically, the distance between the rotor and stator is allowed to vary, but the angle of the rotor is fixed in any one calculation. The stiffness of the bearing and the equilibrium tilt angle were calculated by displacing the rotor vertically or tilting the rotor while adjusting the liquid-air interface to minimize energy. The tilt with the minimum energy was recorded as the equilibrium tilt, while the curvature of the energy versus height or tilt was used to calculate the stiffness. An example of a generated liquid surface mesh can be seen in Fig.~\ref{fig:compare}. 

Not all combinations of constraints resulted in physically realizable bearing configurations. In the model, a configuration was considered unstable if either the center drop separated or the contact of the ring of water on the upper surface vanished at any point during the calculation. For a range of axle and ring volumes, the model was used to determine if the configuration would be stable and, if stable, the equilibrium rotor height, tilt, and tilt stiffness.

\section{Fabrication}

To verify the validity of the model, the bearing design was also experimentally tested. We fabricated bearings with the same parameters as the model starting with $99.99$\% pure aluminum and using single point diamond turning (SPDT) to cut the aluminum to a flat surface ($\textless 10$~nm RMS roughness). Where needed, surfaces were made superhydrophobic by a modified porous aluminum oxide (PAA) growth technique \cite{critpotan} and fluoropolymeric coating composed of spin coated hexamethlydisalizane (HMDS)\cite{HMDSvap} and Solvey Plastics Hyflon AD-60\cite{hyrough}, as reported previously.\cite{HPdura} The stator ring and axle hydrophilic areas were patterned by coating a superhydrophobic surface with photoresist and sputter-depositing Ti, then performing lift off, to form a $1.25$~mm radius circular hydrophilic region for the axle and an annular hydrophilic region with an average $3.35$~mm radius and a width of $1.4$~mm. SPDT was then used to cut out $1$~cm diameter rotors from  $0.5$~mm thick discs with a superhydrophobic surface produced by the same method. SPTD was also used to remove the superhydrophobic structure in a $1.25$~mm radius circle at the center of the rotor to create a matching hydrophilic circle for the axle water to wet. The rotor hydrophilic area was defined by SPDT instead of photolithography to guarantee concentricity of the rotor edge and the axle to within the accuracy of the ultra-precision lathe used for SPDT ($\textless 100$~nm). A rotor and stator pair are shown in Fig.~\ref{fig:bearing}. 

\section{Experimental}

\begin{figure*}[tbh!]
  \centering
	\includegraphics[width=.95\textwidth]{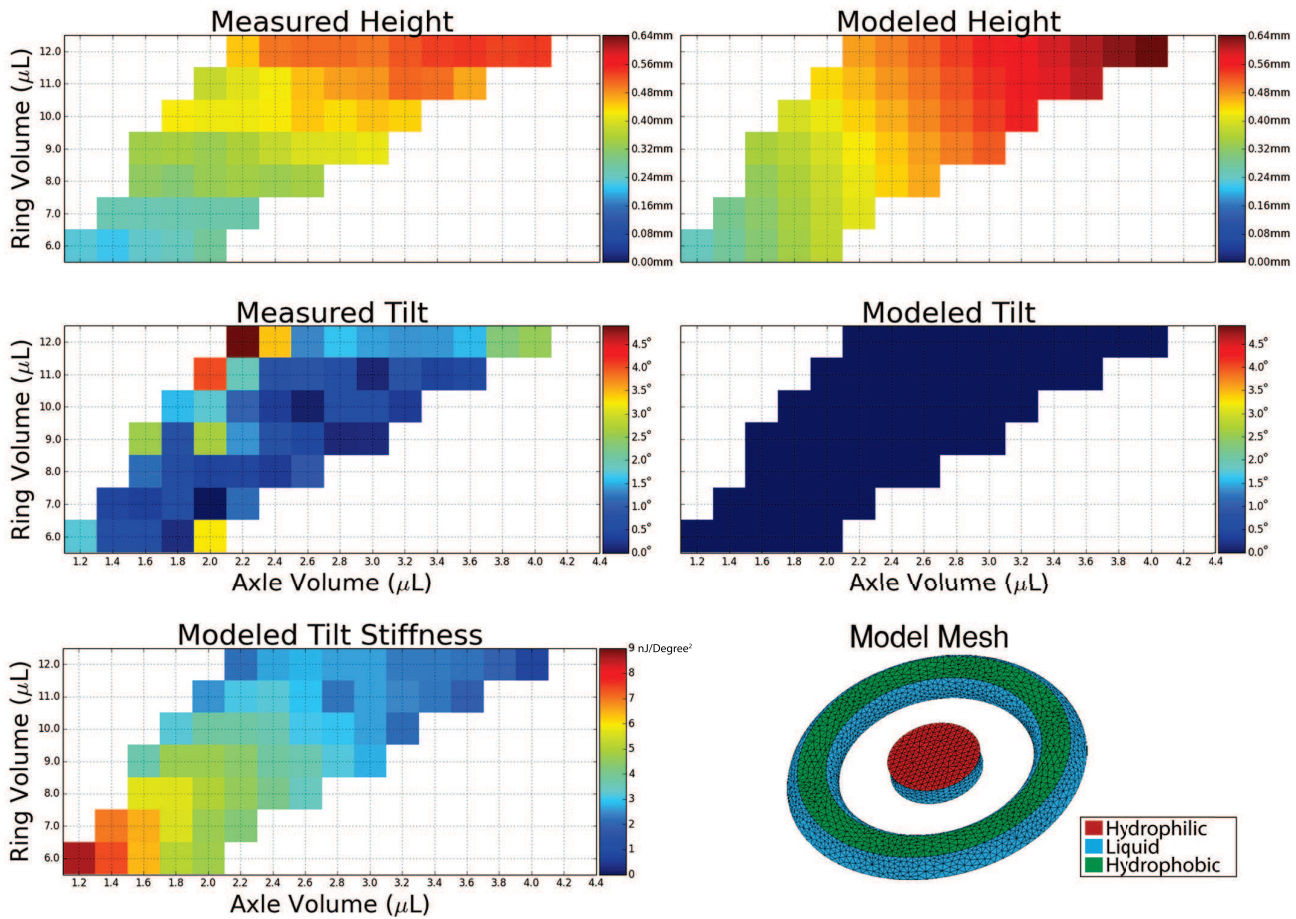}
  \caption{Graphs of height and tilt angle for the examined ring and axle volumes. The stable regions match between the experiments and model, although particular tilt angle is not well correlated. We also show modeled tilt stiffness and an example of our model mesh, which directly simulates only the liquid surface (solid contacts are represented by constraints).}
  \label{fig:compare}
\end{figure*}

Experimentally, the bearings were photographed in profile over a range of center and ring liquid volumes (examples shown in Fig.~\ref{fig:failure}). The parts were allowed to completely dry between tests to minimize volume uncertainties, and the ring and axle hydrophilic regions were filled with water just before placement to minimize evaporation. To achieve full wetting of both axle pads reliably with such small volumes of water, half of the volume had to be placed on each side. Rotors were handled by vacuum tweezers from the back side to ensure that the superhydrophobic surfaces were not damaged. After placement of the rotor, the entire bearing was carefully rotated to ensure that the image was taken in profile perpendicular to the tilt. Image analysis was then used to find the height and angle of tilt. 

In both experiments and models, we find that there are several possible sources of instability. Fig.~\ref{fig:compare} shows the results of experiments and models with the same range of axle and ring volumes. If the axle has insufficient liquid relative to the ring, the liquid on the rotor and stator will not join to form an axle or will re-separate due to the lift from the ring with most of the water on one side, and the rotor will not remain centered (Fig.~\ref{fig:failure}a); this leads to the loss of stability seen in the lower right area in each plot in Fig.~\ref{fig:compare}. If the liquid in the ring is insufficient relative to the axle, then the rotor is lifted partially off of the ring and tips to one side, since balancing on the center drop alone is unstable (Fig.~\ref{fig:failure}b); this leads to the loss of stability seen in the upper left area in each graph in Fig.~\ref{fig:compare}. Finally, if the ring has too much water in it, it bulges asymmetrically, again resulting in a tilted (and often off-center) rotor (Fig.~\ref{fig:failure}c); this requires a volume in the rings such that they have more than a semi-circular crossection, which is larger than the maximum ring volume shown in Fig.~\ref{fig:compare}.

There is an exact match between the predicted and measured regions of stability. This shows that a simple model can predict the stable range of parameters for the bearing. The experimentally measured and numerically modeled rotor heights show reasonable agreement. Both show increasing height of the rotor with increasing volume in either the axle or the ring. Quantitatively, the results agree to within $30$\% with no free parameters in the model. The results for the rotor tilt show less agreement between the experiments and model. The model predicts effectively no tilt in the stable region, while experimentally measurable tilt is observed, particularly near the high ring volume and low axle volume boundary of the stable region. This experimental tilt seems to fluctuate without any clear trend, and is not highly repeatable. It is likely that experimentally this results from the assembly process of the bearing, where the bearing is not guaranteed to start off with zero tilt; i.e. near the edge of the stable region, there may be some hysteresis in the tilt.

\section{Discussion}

With some evidence that the model is accurate, we can then predict the stiffness of the bearing with respect to tilt (vertical stiffness is not generally difficult to achieve, and thus is not analyzed). The results, calculated from the dependence of the energy of the system on tilt, are shown in Fig.~\ref{fig:compare}. Using the same calculation method, for bearings with the same thickness (0.38~mm), a model drop bearing (Fig~\ref{fig:bearing_types}a) has a spring constant of 0.32~nJ/degree$^2$ (drop covers the same area as the full ring and axle bearing), a ring bearing (Fig.~\ref{fig:bearing_types}b) has a spring constant of 0.34~nJ/degree$^2$ (compared to the full ring and axle bearing, the simulated ring bearing has an inner radius equal to the radius of the axle and an outer radius equal to the outer radius of the ring), and the new ring and axle bearing (Fig.~\ref{fig:bearing_types}d) has a spring constant of 2.55~nJ/degree$^2$, which represents a significant improvement. The tilt stiffness of ring and axle bearings is calculated to generally increase with decreasing volume of liquid in either the axle or ring. It is likely that having less liquid in the ring directly increases the stiffness, while less liquid in the axle pulls the rotor closer to the stator and compresses the ring, also increasing the bearing stiffness.

\section{Conclusion}

The surface tension supported ring and axle bearings reported have significant potential for use in MEMS due to their low frictional losses and complete lack of solid on solid contact. These results demonstrate that these devices are stable over a large range of ring and axle volumes, and that this region can be modeled using energy minimization, simplifying design and experimental work with these devices. It also shows that these devices have a higher tilt stiffness for a given size than alternative designs, which is critical since the the tilt stiffness is typically low in surface tension supported bearings. Furthermore, due to the superhydrophobic rotor surface and absence of contact angle hysteresis, ring and axle bearings could have lower drag than previous designs.

A common objection to water-based liquid bearings is that, while promising from a mechanics standpoint, the evaporation of the liquid water makes them of limited practical use. Although the testing presented here was performed with pure water, we have also tested longer term use of a saturated water-CaCl$_2$ solution on superhydrophobic surfaces with patterned hydrophilic regions. This solution has similar surface tension to pure water and since CaCl$_2$ is deliquescent, it can actually pull moisture from the air, rather than allowing the water to evaporate. In practice, this solution was stable over a period of at least 2 years under ambient conditions, and shows no signs of evaporation or corrosion of the underlying surface.

\section*{Acknowledgment}

We would like to thank Dr. Kenneth Brakke at the Mathematics Department of Susquehanna University, the creator of Surface Evolver, for his assistance with understanding and optimizing our simulation.


\end{document}